\documentclass[
 aip,
 amsmath,amssymb,
 reprint
]{revtex4-1}

\usepackage{graphicx}
\usepackage{dcolumn}
\usepackage{bm}

\usepackage[utf8]{inputenc}
\usepackage[T1]{fontenc}
\usepackage{mathptmx}
\usepackage{etoolbox}
\usepackage{xcolor}
\usepackage{float}
\usepackage{hyperref}
\usepackage{orcidlink}
\usepackage{xr}

\usepackage{siunitx}
\usepackage{stackengine}
\usepackage{placeins}
\usepackage{subfigure}

\usepackage{ragged2e}

\usepackage{titlesec}

\makeatletter
\def\@email#1#2{
 \endgroup
 \patchcmd{\titleblock@produce}
  {\frontmatter@RRAPformat}
  {\frontmatter@RRAPformat{\produce@RRAP{*#1\href{mailto:#2}{#2}}}\frontmatter@RRAPformat}
  {}{}
}
\makeatother
\begin{document}

\preprint{AIP/123-QED}

\title[Fabrication Optimization of Suspended Stencil Mask Lithography for Multi-Terminal Josephson Junctions]{Fabrication Optimization of Suspended Stencil Mask Lithography for Multi-Terminal Josephson Junctions}

\author{Justus Teller\,\orcidlink{0000-0003-0032-4200}}
\email{j.teller@fz-juelich.de}
\affiliation{Peter Gr\"unberg Institut (PGI-9), Forschungszentrum J\"ulich, 52425 J\"ulich, Germany}
\affiliation{JARA-Fundamentals of Future Information Technology, J\"ulich-Aachen Research Alliance, Forschungszentrum J\"ulich and RWTH Aachen University, 52425 J\"ulich, Germany}

\author{Abdur Rehman Jalil\,\orcidlink{0000-0003-1869-2466}}
\affiliation{Peter Gr\"unberg Institut (PGI-10), Forschungszentrum J\"ulich, 52425 J\"ulich, Germany}

\author{Florian Lentz\,\orcidlink{0000-0002-8716-6446}}
\affiliation{Helmholtz Nano Facility, Forschungszentrum J\"ulich, 52425 J\"ulich, Germany}

\author{Detlev Gr\"utzmacher\,\orcidlink{0000-0001-6290-9672}}
\affiliation{Peter Gr\"unberg Institut (PGI-9), Forschungszentrum J\"ulich, 52425 J\"ulich, Germany}
\affiliation{JARA-Fundamentals of Future Information Technology, J\"ulich-Aachen Research Alliance, Forschungszentrum J\"ulich and RWTH Aachen University, 52425 J\"ulich, Germany}

\author{Thomas Sch\"apers\,\orcidlink{0000-0001-7861-5003}}
\email{th.schaepers@fz-juelich.de}
\affiliation{Peter Gr\"unberg Institut (PGI-9), Forschungszentrum J\"ulich, 52425 J\"ulich, Germany}
\affiliation{JARA-Fundamentals of Future Information Technology, J\"ulich-Aachen Research Alliance, Forschungszentrum J\"ulich and RWTH Aachen University, 52425 J\"ulich, Germany}

\date{\today}

\begin{abstract}
Stencil mask lithography is an advanced technique for fully \textit{in-situ} fabricating Josephson junctions, which is increasingly being used for multi-terminal Josephson junctions. This study provides information on the optimal mask design and mask reliability. For this, 270 mask designs were systematically fabricated and investigated under scanning electron microscope. Reliable statements are made about mask yield, minimal dimensions, and systematic dependencies on the number of superconducting terminals. We find that stencil mask lithography can be used reliably for fabricating multi-terminal Josephson junctions, enabling lateral mask dimensions down to 40$\,$nm on average.
\end{abstract}

\maketitle

\section{\label{sec:Intro}Introduction}
Stencil mask lithography has a rich history dating back to 1969.\cite{alix_convenient_1969, vazquez-mena_resistless_2015} In contrast to resist patterning, stencil masks can be kept at a distance from the substrate, moved during processing, and used multiple times, since they are mechanically stable and self-supporting.\cite{vazquez-mena_resistless_2015} Moreover, the technique is scalable and applied across several fields.\cite{du_stencil_2017} Recent developments include the fabrication of complex 3-dimensional (3D) surfaces and self-aligned superlattices,\cite{zeng_direct_2025} a plasmonic metasurface of 3D inclined structures,\cite{jeong_threedimensional_2024} next-generation biochemical sensors,\cite{ali_stencilbased_2024} strechable conductors,\cite{sun_stretchable_2021} and transmon  qubits.\cite{tsioutsios_freestanding_2020}\\
In 2019, Schüffelgen~\textit{et al.}\cite{schuffelgen_selective_2019} introduced stencil mask lithography for the fabrication of Josephson junctions with normal-conducting weak link. The advantage of this in situ fabrication technique is that all material layers can be deposited by molecular beam epitaxy (MBE) or electron beam evaporation without breaking the ultra-high vacuum. Consequently, the material interfaces are exceptionally clean, free of residual interface atoms or oxidation. In a superconductor-normal conductor-superconductor (SNS) junction, the Josephson supercurrent flows between two closely spaced superconducting electrodes bridged by the normal-conducting weak-link material. The smaller the separation, the larger the superconducting coupling. Using a stencil mask to deposit the superconducting material the distance between the superconducting electrodes is defined by the width of the mask. The stencil mask technique can be used to fabricate any kind of SNS weak link Josephson junction, comprising different weak link materials, such as metals, semiconductors or topological materials. Over the last years, a broad range of research studies has been published using this fabrication technique, including two-terminal Josephson junctions,\cite{schuffelgen_selective_2019, schmitt_integration_2022a, schmitt_anomalous_2022, rosenbach_ballistic_2023, zimmermann_topological_2024, hanna_onchip_2025, jalil_engineering_2025} three-terminal Josephson junctions,\cite{kolzer_supercurrent_2023a, behner_superconductive_2025a} and whole networks of Josephson junctions.\cite{teller_frustrated_2025}\\
\begin{figure}[h]
    \centering
    \includegraphics[width=0.35\textwidth]{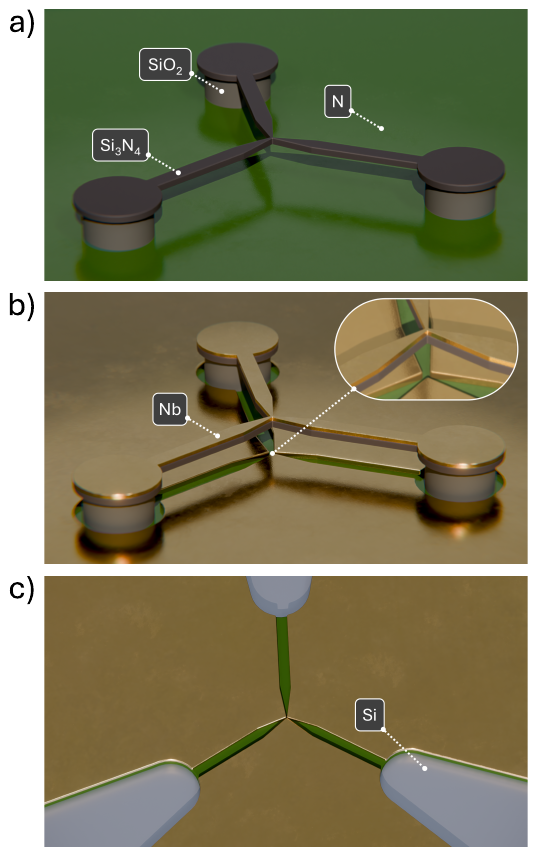}
    \caption{a) Visualisation of a suspended stencil mask for a three-terminal junction with normal conducting material (N) deposited underneath. The mask itself is made of three pillars (SiO$_2$) holding up the top Si$_3$N$_4$. b) Deposited Nb on top of the mask. The mask casts a shadow onto the underlying Si substrate. Therefore, the geometry of the mask is directly connected to the resulting device geometry on the substrate.  c) Finished three-terminal junction after mask removal and subsequent etching step to separate the superconducting electrodes, revealing the Si substrate.}
    \label{fig:Schematic}
\end{figure}
\begin{figure*}[t]
    \centering
    \includegraphics[width=0.9\textwidth]{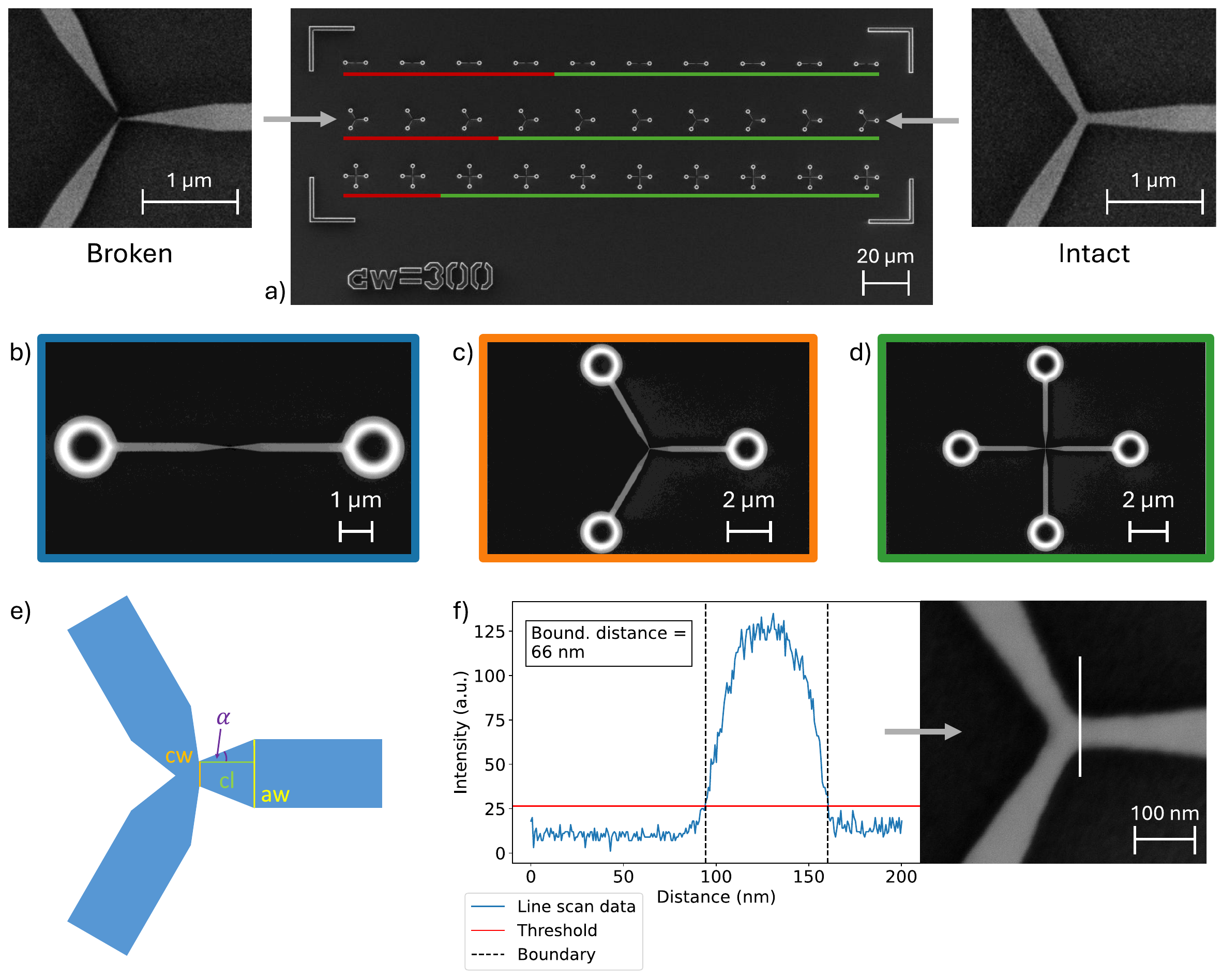}
    \caption{Fabricated masks and determination of mask dimensions. a) Scanning electron microscopy image of a fabricated design cell containing three rows for two-, three-, and four-terminal masks, respectively. The arm width was set for each design cell (here $aw = 300\,$nm) and the center width was increased from left (10$\,$nm) to right (100$\,$nm) in 10-nm-steps. Broken (left) and intact (right) stencil masks were easily identifiable under SEM. The red-marked masks were broken and the green-marked masks were intact for this particular design cell. b)-d) In total masks for two-, three-, and four- terminal Josephson junctions were investigated. The images correspond to the masks in a) and their respective frame color equals that of the data, which are presented throughout the study, for the corresponding number of arms. e) Geometry of a three-terminal junction and design parameters used in this study: center width ($cw$), center length ($cl$), arm width ($aw$). Together, they define the angle of sharpness ($\alpha$). f) An exemplary center width measurement of the mask in c) under SEM using a line scan. The position of the line scan is indicated in the SEM image on the right as a white line. In this particular example, the center width was determined to be $cw = 66\,$nm. Information on determining the threshold value (red line) is given in Supplemental Material.}
    \label{fig:Experiment}
\end{figure*}
In general, more complex Josephson junctions, i.e., multi-terminal junctions, have attracted increasing attention recently.\cite{pfeffer_subgap_2014, riwar_multiterminal_2016, draelos_supercurrent_2019a, graziano_transport_2020, pankratova_multiterminal_2020, coraiola_2023, coraiola_2024} They are defined as a Josephson junction with more than two superconducting contacts that all share the same weak link.\cite{duhot_cross_2009a, riwar_multiterminal_2016, pankratova_multiterminal_2020}
For multi-terminal Josephson junctions one can distinguish between two model descriptions: In the frist model, all superconducting weak link channels meet at a  central scattering region in the middle\cite{duhot_cross_2009a, riwar_multiterminal_2016, pankratova_multiterminal_2020, graziano_selective_2022, coraiola_2023, matute_2024} (central scattering transport model)  and, in the second model, each superconducting contact pairwise shares transport channels with a neighbouring one in a network.\cite{draelos_supercurrent_2019a, graziano_transport_2020, arnault_multiterminal_2021a, arnault_dynamical_2022} The latter can be described by a resistively and capacitively shunted junction (RCSJ) network using Kirchhoff's circuit laws.\cite{graziano_transport_2020, arnault_dynamical_2022} The central scattering model is of great interest because it allows direct coupling of multiple electrodes, thereby enabling novel device functionalities.\cite{matute_2024} Furthermore, it is predicted that artificial topological states can be created, e.g. topological Weyl singularities.\cite{riwar_multiterminal_2016,peralta-gavensky_2023} However, achieving the central scattering region model poses particular challenges for the junction geometry, i.e. all electrodes should be as close as possible in the center part of the device. 

This study systematically investigates the optimal geometric fabrication parameters for stencil masks of two-, three-, and four-terminal Josephson junctions.  Figure~\ref{fig:Schematic} a) illustrates the layout of a stencil mask for a three-terminal Josephson junction. The suspended Si$_3$N$_4$ mask rests on three SiO$_2$ pillars. The weak-link material of the junction, e.g., a metal, semiconductor, or topological material, is present under the bridge. One way to define the weak-link geometry in situ is to employ selective-area growth, as described by Sch\"uffelgen \textit{et al.}\cite{schuffelgen_selective_2019} When the superconductor layer is deposited at a fixed angle from above, it casts a shadow corresponding to the mask layout [cf. Fig.~\ref{fig:Schematic} b)]. This defines the separation of the superconducting electrodes. The device fabrication is finished by a etching of the superconductor layer to define separated electrodes, which are later contacted [cf. Fig.~\ref{fig:Schematic} c)]. As shown in the inset of Fig.~\ref{fig:Schematic}~b), the closest distance is achieved in the central region of the mask, where the arms of the mask have the smallest width. A Josephson supercurrent is expected to flow here. Mask optimization occurs within a trade-off context: the wider a stencil mask is, the more stable it is. However, the resulting coupling via the weak link of the Josephson junction is becoming smaller. To achieve good superconducting coupling, we therefore focus on fabricating the masks as narrow as possible but stable at the same time. We will also discuss how to ensure that the fabrication process leaves the mask intact. The masks are designed in a way that each contact is closest to every other contact in the center of the weak link, making the central scattering region transport model more likely.

For a systematic study on the design limits of stencil masks, we created 270 different mask designs, each one fabricated a total of four times on two different samples. We then investigated by scanning electron microscopy (SEM) which of the masks were broken or intact. For the narrowest intact masks, we looked at how their geometrical dimensions were realized compared to their designed dimensions.\\
\par

\section{\label{sec:Fabrication}Mask Fabrication}
The mask was fabricated using a Si(111) substrate covered with a 300-nm-thick SiO$_2$ layer and a 100-nm-thick Si$_3$N$_4$ layer by plasma-enhanced chemical vapor deposition and a subsequent heating to 1000$\,^\circ$C (5\;min). For patterning the mask, the substrate was cleaned with acetone (5$\,$min) and isopropanol (5$\,$min), and then pre-baked at 130$\,^\circ$C (5$\,$min). Then Medusa photoresist (\mbox{Medusa 82:AR 600-07, ratio 1:2}) was spin coated on the sample (4000$\,$rpm, 30$\,$sec) and soft baked at 150$\,^\circ$C (10$\,$min).\\
Afterwards, the pattern was written by electron beam lithography (EBL). The gaussian beam EBL tool Raith EBPG 5200 with an acceleration voltage of 100 kV was used. The design was split into fine and coarse parts which were written with beam currents of 500 pA and 150 nA and beam step sizes of 1 nm and 50 nm, respectively. A multipass scanning mode with four passes was applied for the fine structures while the coarse structures were written in a single pass. The total base dose in both cases amounted to 7096 µC/cm² with an additional proximity effect correction with dose factors ranging from 1.0 up to 1.63.\\
Subsequently, the sample was post exposure baked at 170$\,^\circ$C (10$\,$min) and the resist was developed with AR-300-44 (80$\,$sec). 
\begin{figure}
    \centering
    \includegraphics[width=0.35\textwidth]{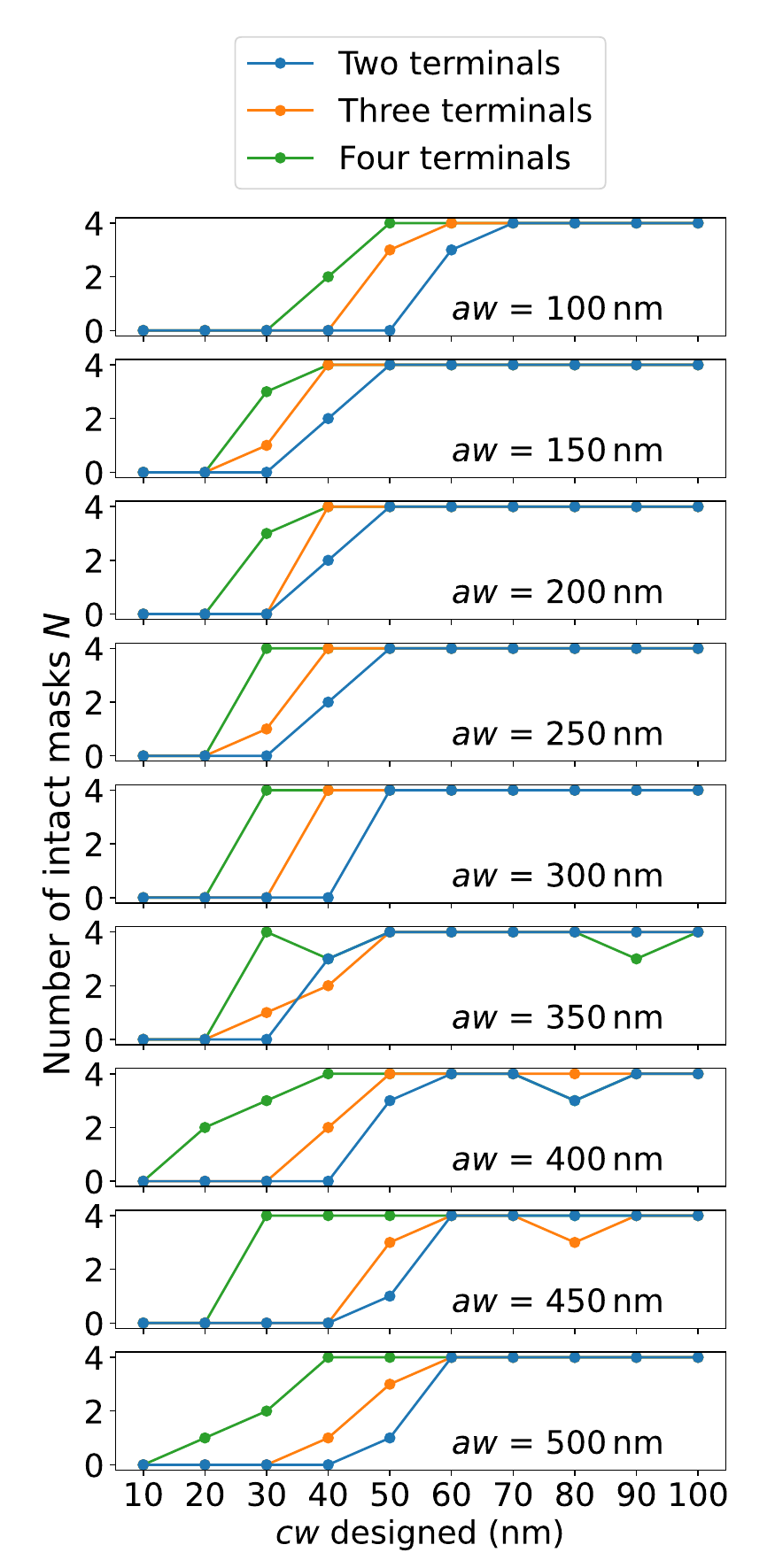}
    \caption{Number of intact masks $N$ depending on the designed center width $cw$ and designed arm width $aw$ with an accuracy of 25$\,\%$. The more arms a mask has, the smaller its center width can be designed. Above a certain designed $cw$, the number of intact masks is close to 100$\,\%$.}
    \label{fig:standing_prob}
\end{figure}
The written pattern was then etched into the Si$_3$N$_4$ layer by inductively-coupled plasma reactive-ion etching (ICP-RIE) (RF power: 25$\,$W, ICP power: 100$\,$W, CHF$_3$: 55$\,$sccm, O$_2$: 5$\,$sccm, time: 2min$\,$55sec). Then the Si$_3$N$_4$ layer was under etched by buffered hydrofluoric acid (BOE) (5$\,$min) and 1$\,\%$-concentrated hydrofluoric acid (3$\,$min). The acid also removed the Medusa photoresist, so the wet-etch step served as a cleaning step as well.

\begin{figure*}[t!]
    \centering
    \includegraphics[width=0.8\textwidth]{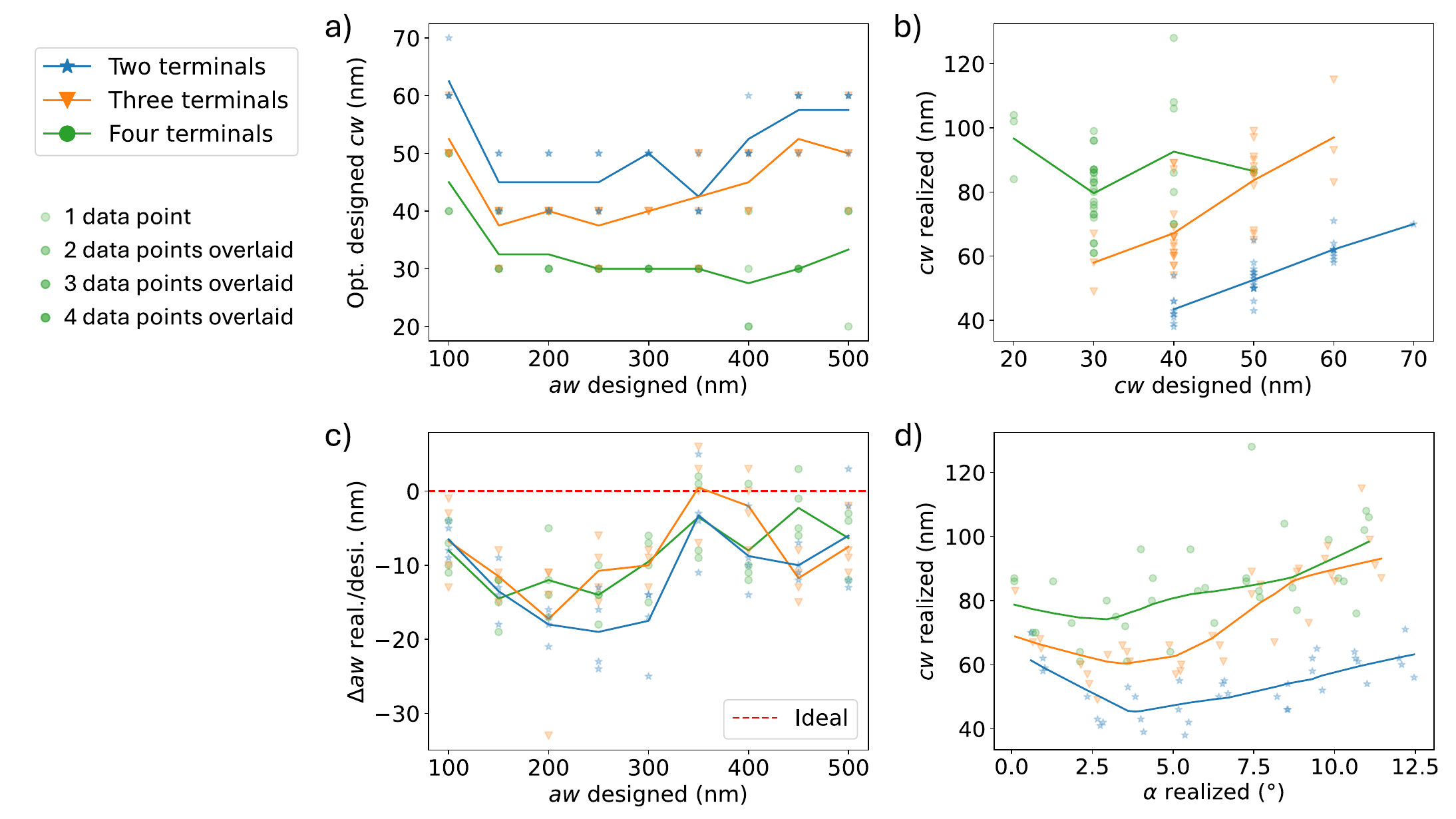}
    \caption{Geometrical results from the SEM line scans. Each data point is indicated by a marker corresponding to the number of terminals of its mask. The solid lines serve as a guide to the eye\footnote{In Figs.~a)-c), the solid lines connect the means of the $y$-values scattering at a certain $x$-value. In Figs.~d)~and~e), the solid lines are determined via locally weighted scatterplot smoothing (LOWESS). See Supplementary Material for more details.} and data points are transparent. a) The optimally designed $cw$ depending on the designed $aw$. On average, four-terminal masks can be designed down to 30$\,$nm, three-terminal masks masks down to 40$\,$nm, and two-terminal masks down to 50$\,$nm. b) The smallest realized center width $cw$ vs. designed center width. Two- and three-terminal masks show a linear dependence on $cw$ design whereas the realized $cw$ of the four-terminal masks scatter around a constant value. c) The difference between realized and designed arm width. The dashed red line indicates the case of matching of designed and achieved values of $aw$. All values scatter equally around $-10\,$nm. d) Relation between the realized angle of sharpness $\alpha$ and the realized center width $cw$, assuming an as-designed center length of $cl = 1\,\mu$m.}
    \label{fig:Experimental_Results}
\end{figure*}

For this study, we created 270 different mask designs using the PHIDL software.\cite{mccaughan_phidl_2021} All design parameter are given in Fig.~\ref{fig:Experiment}~e). The designs iteratively change the center width ($cw$) and the arm width ($aw$), whereas the center length ($cl$) stays constant at 1$\,\mu$m. This results in a change of the angle of sharpness, $\alpha = \text{arctan}\left[\left(aw-cw\right)/2cl\right]$. 
The larger $\alpha$ is the more the superconducting coupling is located in the center part, because the coupling is stronger the smaller the electrode separation is. On a sample, the designs are ordered in so-called design cells, each having designs with constant arm width [see Fig.~\ref{fig:Experiment}~a)]. The arm widths range from 100$\,$nm to 500$\,$nm in 50$\,$nm-steps. Within the design cell, there are three rows with designs for two-, three-, and four-terminal masks, respectively. In each row, the center width is increased from 10$\,$nm to 100$\,$nm in 10$\,$nm-steps (from left to right).

\section{\label{sec:Experimental}Experimental Results}

In order to evaluate systematically the mask properties all fabricated masks were inspected by scanning electron microscopy. From every mask, an image was taken on which it is visible whether the mask is broken or intact [see Fig.~\ref{fig:Experiment}~a)]. Starting with the first intact mask in each row (from left to right), the geometric parameters $cw$ and $aw$ were measured using line scans. [see Fig.~\ref{fig:Experiment}~f) and Supplementary Material]. These values are used for the discussion in this study.

As mentioned above, the weakest point regarding the mask stability is its narrow middle part. Therefore, it is expedient to look at the probability of mask integrity depending on the center width design parameter $cw$. In general, the center width should be as small as possible to ensure strong Josephson coupling. In Fig.~\ref{fig:standing_prob}, the number of intact masks $N$ as a function of the designed center width is given for arm widths $aw$ ranging from 100\,nm to 500\,nm, i.e. for each design cell. A strong dependence between number of intact masks and center width is evident. One finds that above a certain designed center width, it is possible to fabricate the masks reliably with close to $100\,\%$ probability. The more terminals a mask has, the smaller its $cw$ can be designed. This is summarized in Fig.~\ref{fig:Experimental_Results}~a) where, for each arm width $aw$, the optimal (i.e. first intact mask from left to right) $cw$ design is indicated. For the given fixed fabrication parameters for all types of masks, the two-terminal masks can be designed down to a $cw$ of $50\,$nm, three-terminal masks down to $40\,$nm, and four-terminal masks down to 30$\,$nm.

Since we are interested in the smallest possible dimensions, we will analyze and compare for each type of mask the smallest intact masks we find in the particular rows of the design cells. In the following, the actual geometries of these masks are discussed in detail. In Fig.~\ref{fig:Experimental_Results}~b) the smallest achieved center widths for the three types of masks are plotted as a function of the designed center widths. We find that contrary to the design limits outlined above, a two-terminal mask can be realized with the smallest center width, whereas the smallest realized $cw$ increases the more terminals a mask has. For two-terminal masks $cw$ values as small as 40\,nm are obtained. Furthermore, in contrast to the two-terminal masks the smallest achieved $cw$ of the three- and four-terminal masks exceeds the designed values. As indicated by the blue and orange lines in Fig.~\ref{fig:Experimental_Results}~b), two- and three-terminal masks show quasi-linear behavior between realized and designed $cw$. In contrast, the values of the four-terminal masks scatter around a constant value of $90\,$nm on average. The reason for both dependencies (contrast between $cw$ design limit and realization limit and linear/constant design dependence) is likely the angle between the mask arms, i.e. 180$^\circ$ for two-terminals, 120$^\circ$ for three-terminals, and 90$^\circ$ for four-terminals. Electron beam lithography and wet-etching are both known to have angle dependent fabrication effects: the electron beam lithography is subject to pattern specific line width variations by the proximity effect\cite{ren_proximity_2004, zhou_electron_2005} and the chemical wet etch rate can be limited by the transport of reactant species to the etched surface as well as the transport of end products away from the surface.\cite{sarangan_nanofabrication_2017} These transports are more limited in 90$^\circ$ angles than at straight lines. Although BOE etches the SiO$_2$ at a significantly higher rate, the Si$_3$N$_4$ is etched by it as well.

Beside the center width value $cw$, which determines the closest distance between the superconducting electrodes and by that the strength of the superconducting coupling the angle of sharpness $\alpha$ is of equal importance for multi-terminal junctions. It expresses how much the superconducting coupling is restricted to the center part. The larger $\alpha$ is, the more the superconducting electrode are separated when leaving the junction center. Figure~\ref{fig:Experimental_Results}~d) shows the relationship between the angle of sharpness, denoted by $\alpha$, and the minimal width of the center, denoted by $cw$. The minimum center width increases a little bit with increasing $\alpha$, which can be explained by the proximity effect during electron beam lithography. We routinely realized an angle of sharpness of more than 10$^\circ$, which is advantageous for obtaining multi-terminal junctions operated in the central scattering region transport model. 

\section{\label{sec:Conclusion}Conclusion}
We demonstrated that stencil masks for in situ Josephson junction fabrication can be fabricated very reliable with close to $100\,\%$ probability. Hereby, the masks can be fabricated as small in width as 40$\,$nm. This study should serve as a basis for complex future projects. It gives insights into how to design stencil masks more optimally. However, this study is limited to the fabrication of the stencil masks themselves, material deposition was neglected. Once a material is deposited on the masks, the additional strain might break the most unstable ones. Therefore, it is advisable to leave some room for failure in future mask designs. Additionally, there is the option to use an entrenched mask instead of a suspended one (e.g. cf. Ref.~[\onlinecite{jalil_engineering_2025}]). Other mask geometries will likely follow the principles outlined in this work.

In future, one could use the PHIDL software \cite{mccaughan_phidl_2021} in connection with the results of this study to create a library of reliable masks to choose from for projects. This work will enable other researchers to fabricate Josephson junctions with extremely small weak link lengths using the stencil mask lithography technique. In general, stencil masks work reliable for short weak links as well as for multi-terminal Josephson junctions.

\begin{acknowledgments}
We thank Herbert Kertz for technical assistance and Sebastian Droege for the stencil mask visualisations (Fig.~\ref{fig:Schematic}). All samples have been prepared at the Helmholtz Nano Facility.\cite{albrecht_hnf_2017} This work is funded by the Deutsche Forschungsgemeinschaft (DFG, German Research Foundation) under Germany's Excellence Strategy – Cluster of Excellence Matter and Light for Quantum Computing (ML4Q) EXC 2004/1 – 390534769, by the German Federal Ministry of Education and Research (BMBF) via the Quantum Futur project ‘MajoranaChips’ (Grant No. 13N15264), as well as the Bavarian Ministry of Economic Affairs, Regional Development and Energy within Bavaria’s High-Tech Agenda Project ”Bausteine für das Quantencomputing auf Basis topologischer Materialien mit experimentellen und theoretischen Ansätzen” (grant no. 07 02/686 58/1/21 1/22 2/23).\\
\end{acknowledgments}

\textit{Data availability} --- The data that support the findings of this article are openly available.~\cite{data_ref}

\clearpage
\widetext

\titleformat{\section}[hang]{\bfseries}{\MakeUppercase{Supplemental Note} \thesection:\ }{0pt}{\MakeUppercase}
\setcounter{section}{0}
\setcounter{equation}{0}
\setcounter{figure}{0}
\setcounter{table}{0}
\setcounter{page}{1}
\renewcommand{\thesection}{\arabic{section}}
\renewcommand{\thesubsection}{\Alph{subsection}}
\renewcommand{\theequation}{S\arabic{equation}}
\renewcommand{\thefigure}{S\arabic{figure}}
\renewcommand{\figurename}{Supplemental Figure}
\renewcommand{\tablename}{Supplemental Table}
\renewcommand{\bibnumfmt}[1]{[S#1]}
\renewcommand{\citenumfont}[1]{S#1}

\begin{center}
\textbf{\large Fabrication Optimization of Suspended Stencil Mask Lithography for\\ Multi-Terminal Josephson Junctions (Supplemental Material)}
\end{center}

\section{Abstract}
In this supplemental material, we explain the technique used to determine geometrical values from the scanning electron micrographs (Suppl. Sec.~\ref{sec:SUPP_SEM_Measurement}), comment on the methods used to show data in Fig.~\ref{fig:Experimental_Results} and Figs.~\ref{fig:SUPP_Opt_Pos}-\ref{fig:SUPP_cw_real_vs_alpha} (Suppl. Sec.~\ref{sec:SUPP_Fig4}), and discuss the differences between the two samples of this study (Suppl. Sec.~\ref{sec:SUPP_Diff_Samples}).

\section{Measurement Technique}\label{sec:SUPP_SEM_Measurement}
The data for this study is based on measurements with scanning electron microscope (SEM). For Fig.~\ref{fig:standing_prob}, one SEM image was taken of every mask in which it is visible whether the mask is intact or broken. For all other geometrical data, line scans through SEM images were used for measurements [see Fig.~\ref{fig:Experiment}~f)]. To determine a distance within the line scan data, a threshold is first determined where the first data point (left and right of a feature) above it is set as a boundary for the distance measurement. The threshold is determined using the first 20$\,\%$ of a line scan data as background signal, e.g. from 0 to about 40$\,$nm in Fig.~\ref{fig:Experiment}~f). Here, the signal offset is determined by calculating the background mean ($\mu$). The noise level is determined by taking the background standard deviation ($\sigma$). The threshold is set to the signal offset plus five times the noise level (threshold = $\mu + 5\sigma$). The line scan measurements were used to determine the center width ($cw$) and the arm width ($aw$).

\section{Comment about display of data in Fig.~\ref{fig:Experimental_Results} and Figs.~\ref{fig:SUPP_Opt_Pos}-\ref{fig:SUPP_cw_real_vs_alpha}}\label{sec:SUPP_Fig4}
In main text Fig.~\ref{fig:Experimental_Results}, each mask measurement is indicated individually. The markers are transparent for indicating overlays. The solid lines serve as a guide to the eye where each line stands for the behavior of two- (blue), three- (orange), or four-terminal (green) masks. To determine the solid lines accurately, two methods were used:
\begin{itemize}
    \item For the data in Figs.~\ref{fig:Experimental_Results}~a)-c),~\ref{fig:SUPP_Opt_Pos},~\ref{fig:SUPP_cw_real_vs_designed},~and~\ref{fig:SUPP_aw_real_vs_designed}, the y-values scatter at a specific x-value. The solid line connects the means of the y-values scattering at a certain x-value.
    \item For the data in Figs.~\ref{fig:Experimental_Results}~d),~\ref{fig:SUPP_cw_real_vs_aw_real},~and~\ref{fig:SUPP_cw_real_vs_alpha}, the x-values are more scattered than for the other graphs. Here, the solid line is determined via a locally weighted scatterplot smoothing (LOWESS; fraction of data used for weighting: 0.5).
\end{itemize}

\section{Differences between first and second sample}\label{sec:SUPP_Diff_Samples}
\begin{figure}[h]
\centering
\includegraphics[width=0.98\textwidth]{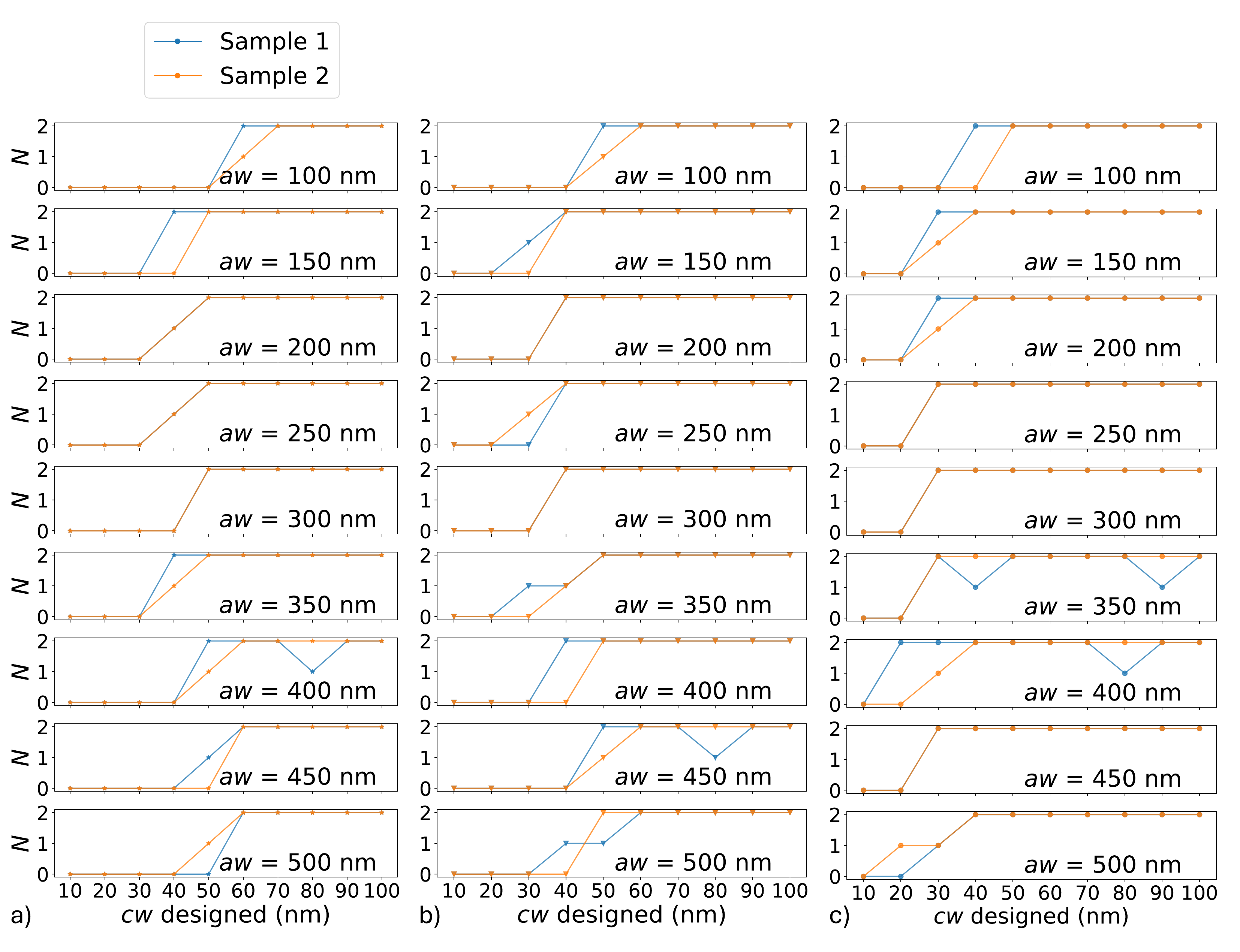}
\caption{Number of intact masks $N$ with two-terminals (a), three-terminals (b), and four-terminals (c). Data shown for each sample individually, with every mask design twice on each sample.}
\label{fig:SUPP_Standing_Prob}
\end{figure}

\begin{figure}[h]
\centering
\subfigure[~Two-terminal masks.]{\includegraphics[width=0.45\textwidth]{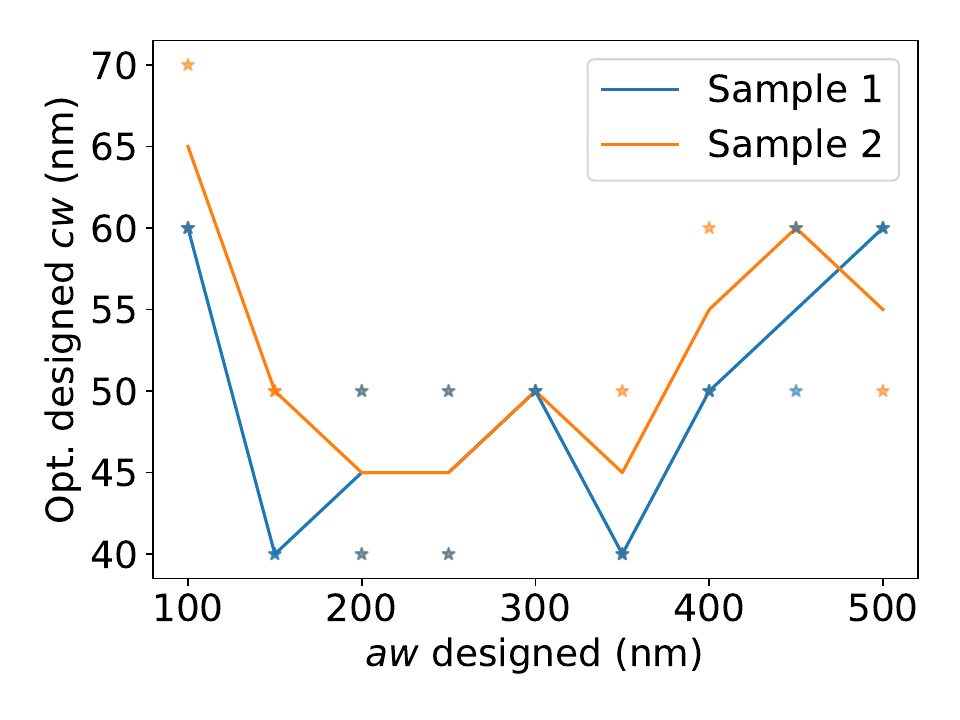}}
\subfigure[~Three-terminal masks.]{\includegraphics[width=0.45\textwidth]{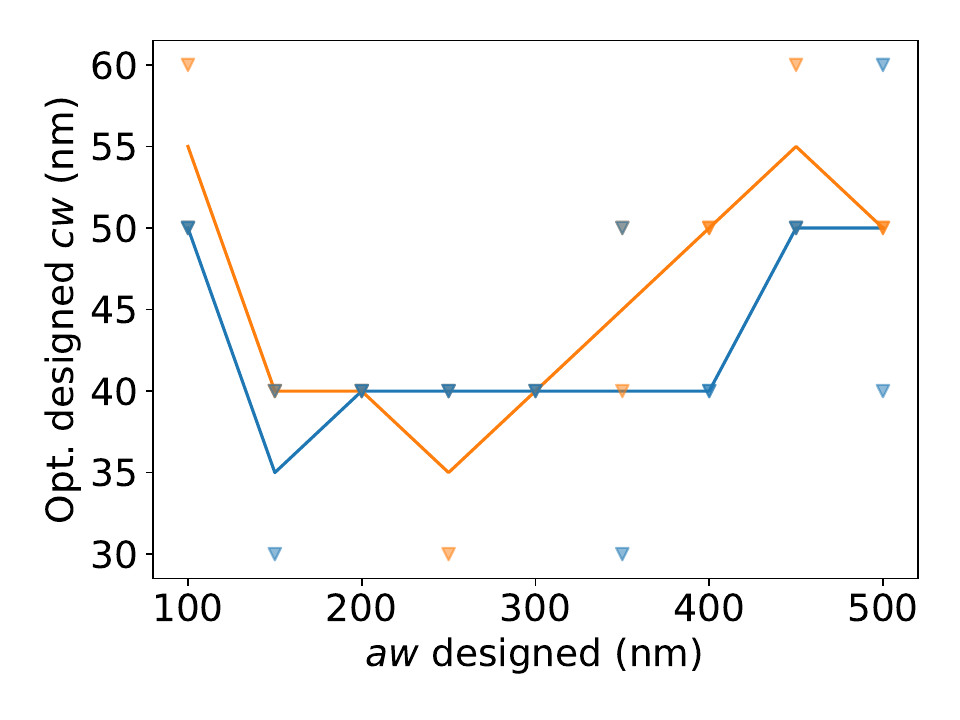}}\\
\subfigure[~Four-terminal masks.]{\includegraphics[width=0.45\textwidth]{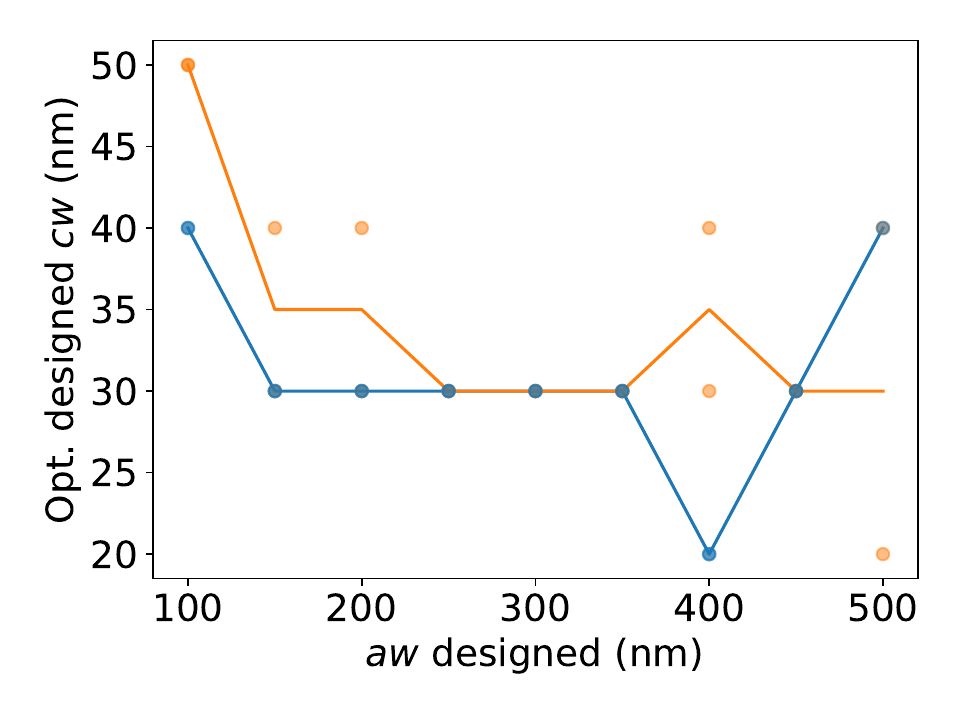}}
\caption{Optimally designed center width depending on the designed arm width. The lines serve as a guide to the eye, as described in Suppl. Sec.~\ref{sec:SUPP_Fig4}.}
\label{fig:SUPP_Opt_Pos}
\end{figure}

\begin{figure}[h]
\centering
\subfigure[~Two-terminal masks.]{\includegraphics[width=0.45\textwidth]{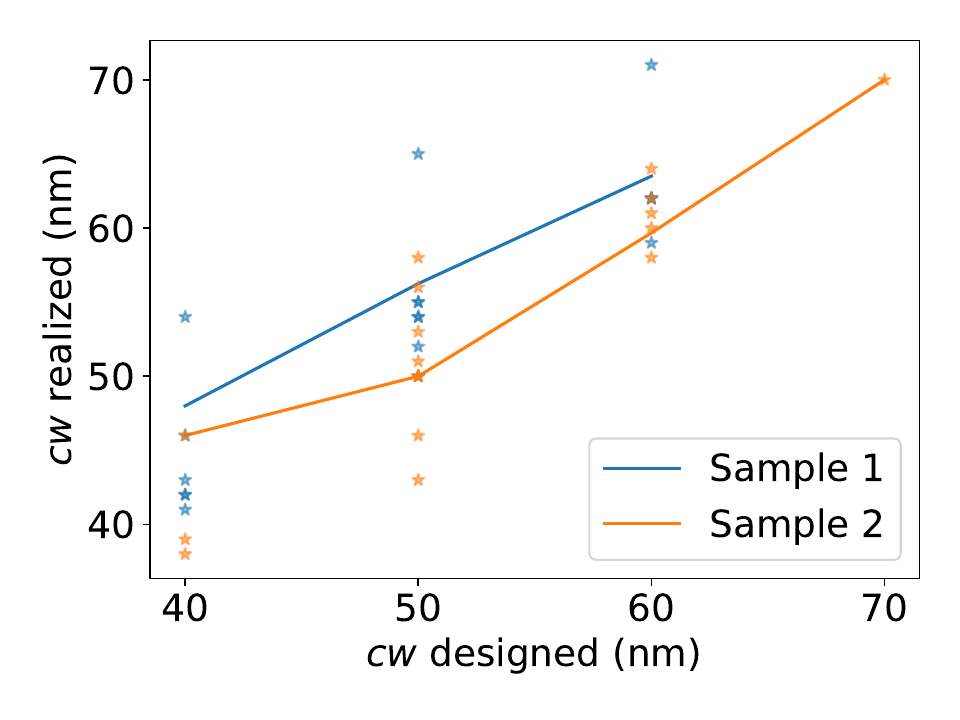}}
\subfigure[~Three-terminal masks.]{\includegraphics[width=0.45\textwidth]{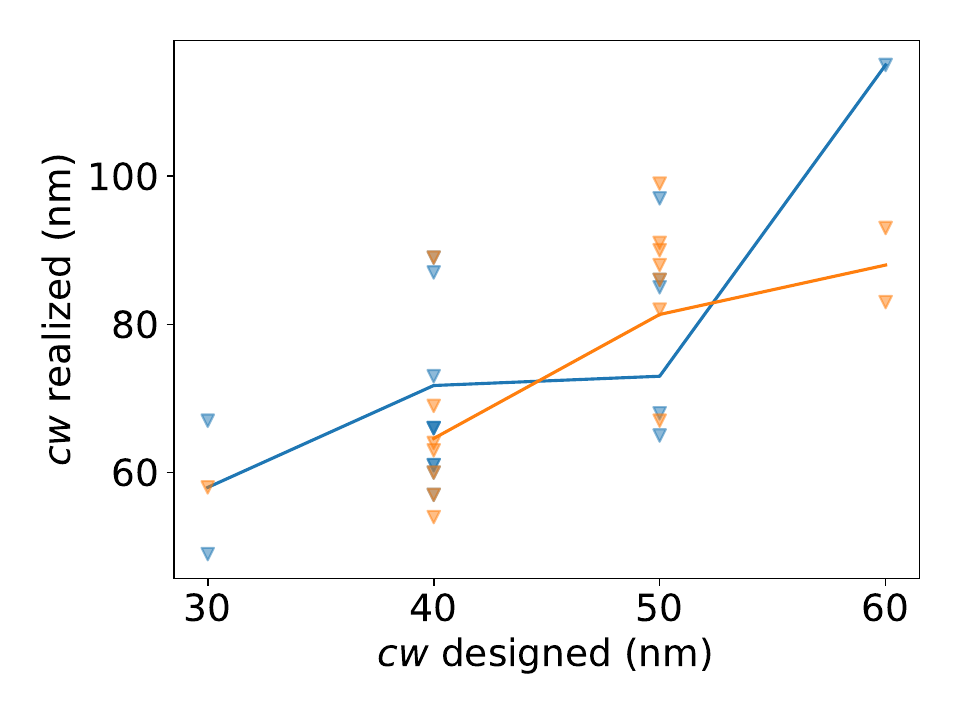}}\\
\subfigure[~Four-terminal masks.]{\includegraphics[width=0.45\textwidth]{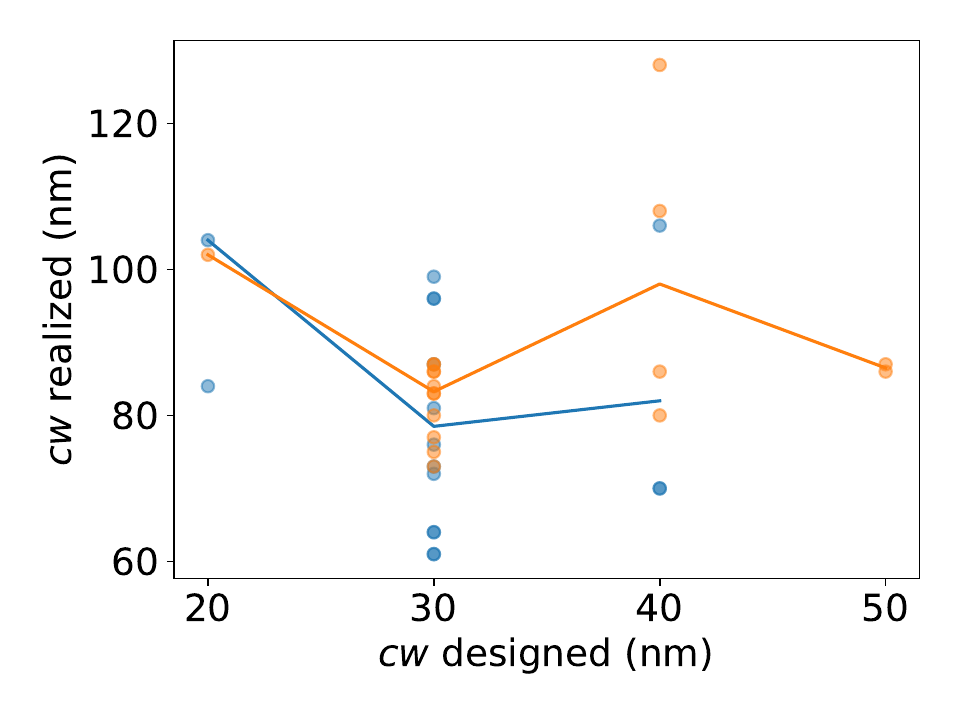}}
\caption{The realized compared to the designed center width. The lines serve as a guide to the eye, as described in Suppl. Sec.~\ref{sec:SUPP_Fig4}.}
\label{fig:SUPP_cw_real_vs_designed}
\end{figure}

\begin{figure}[h]
\centering
\subfigure[~Two-terminal masks.]{\includegraphics[width=0.45\textwidth]{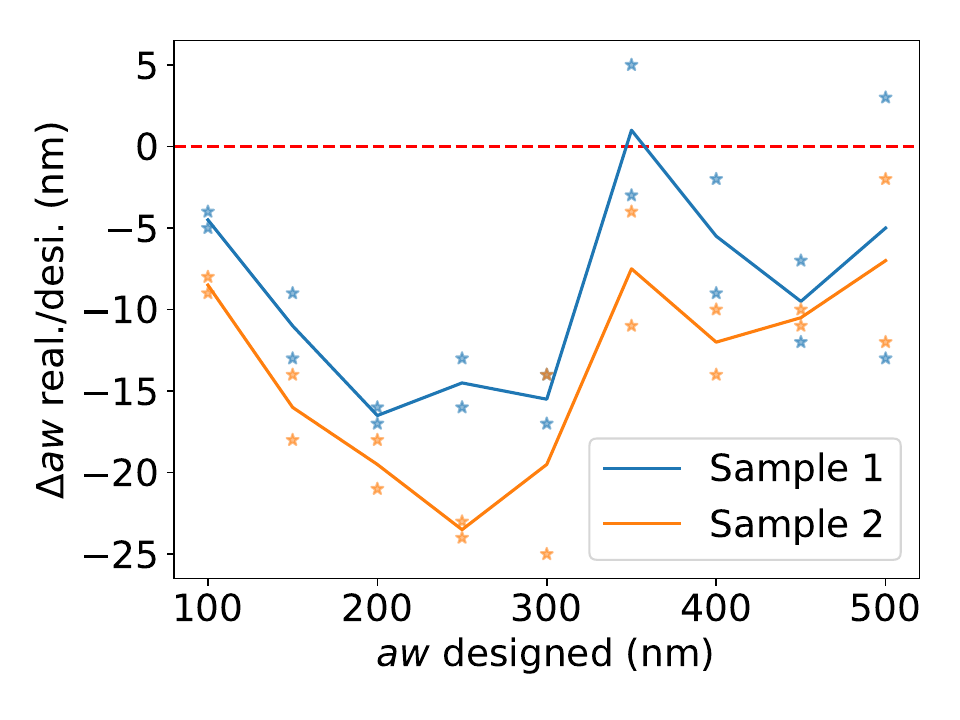}}
\subfigure[~Three-terminal masks.]{\includegraphics[width=0.45\textwidth]{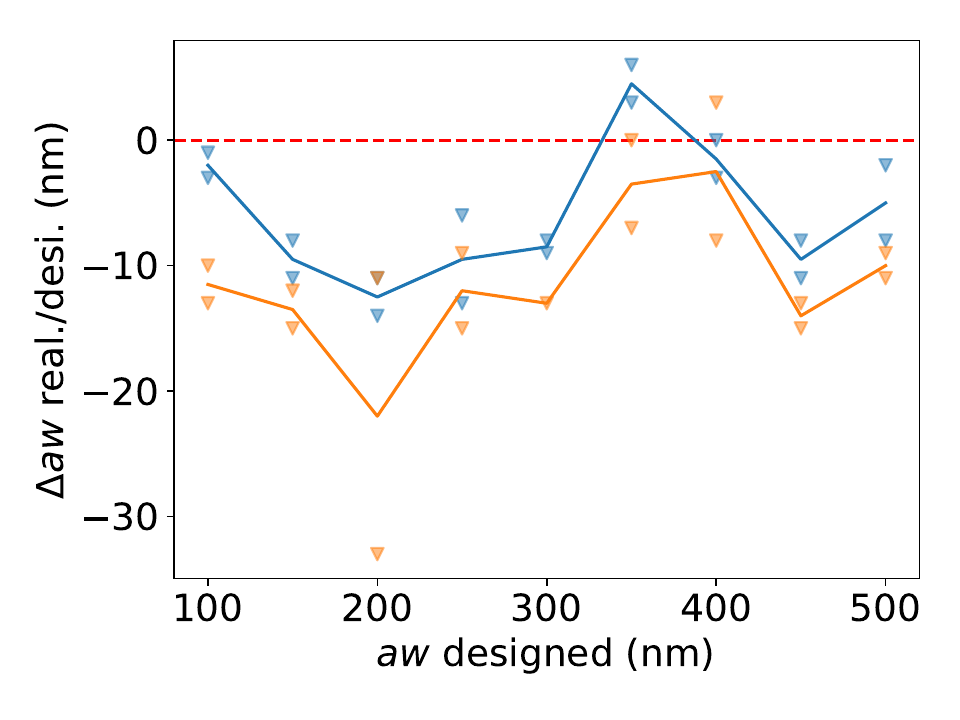}}\\
\subfigure[~Four-terminal masks.]{\includegraphics[width=0.45\textwidth]{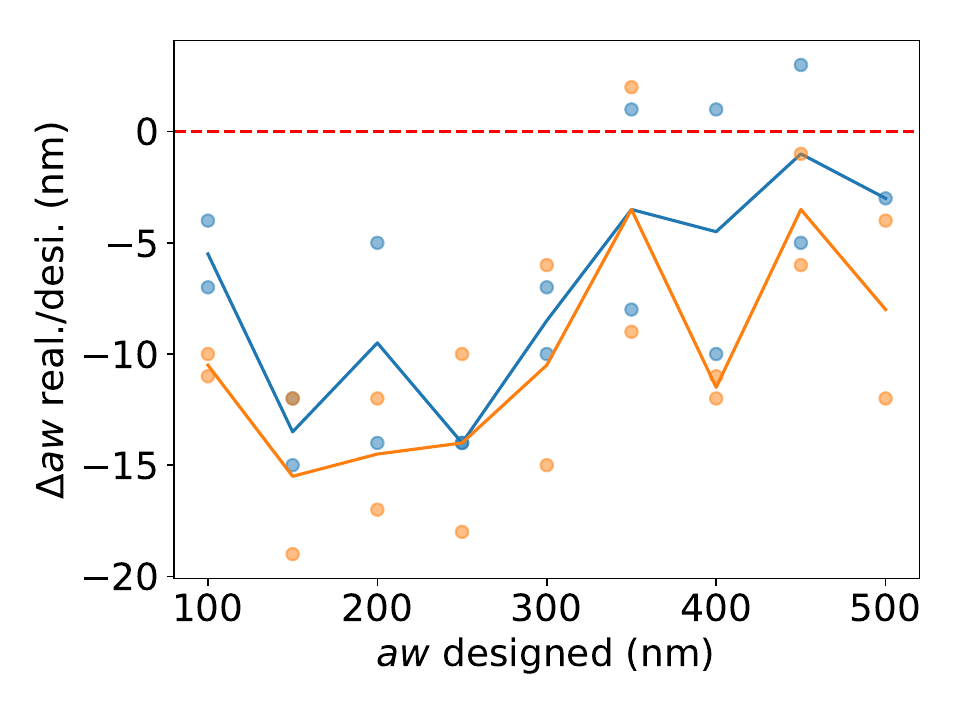}}
\caption{Difference between realized and designed arm width. The lines serve as a guide to the eye, as described in Suppl. Sec.~\ref{sec:SUPP_Fig4}.}
\label{fig:SUPP_aw_real_vs_designed}
\end{figure}

\begin{figure}[h]
\centering
\subfigure[~Two-terminal masks.]{\includegraphics[width=0.45\textwidth]{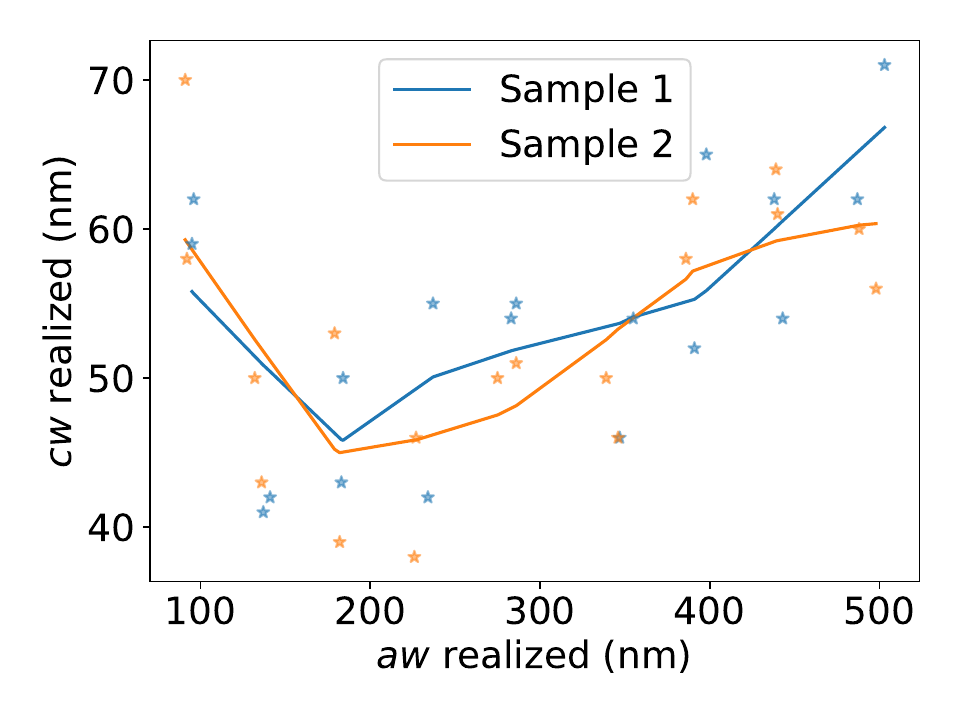}}
\subfigure[~Three-terminal masks.]{\includegraphics[width=0.45\textwidth]{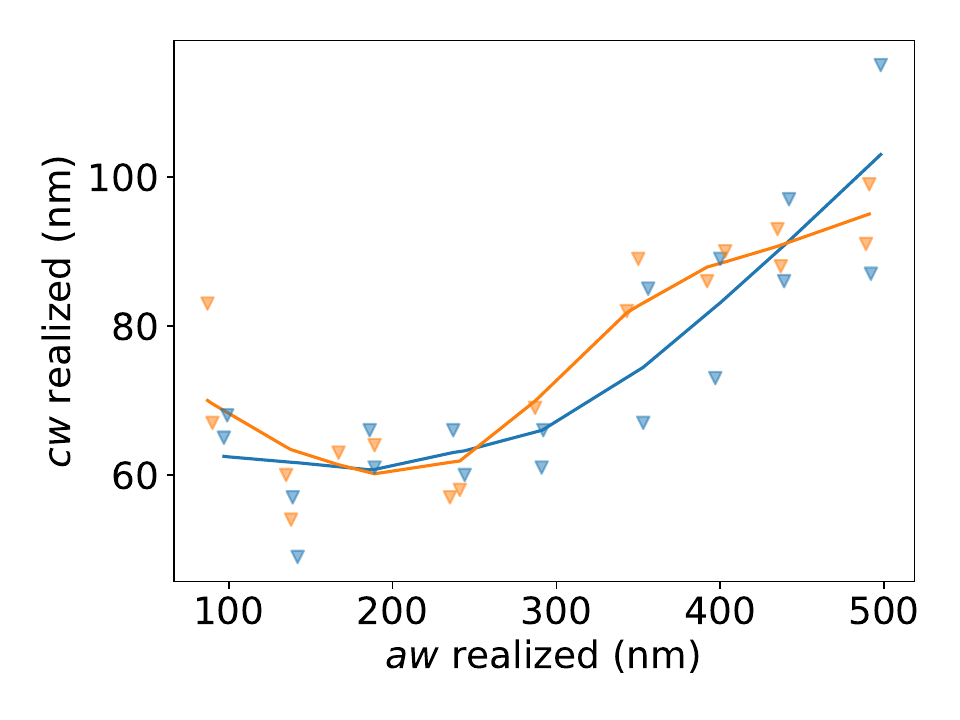}}\\
\subfigure[~Four-terminal masks.]{\includegraphics[width=0.45\textwidth]{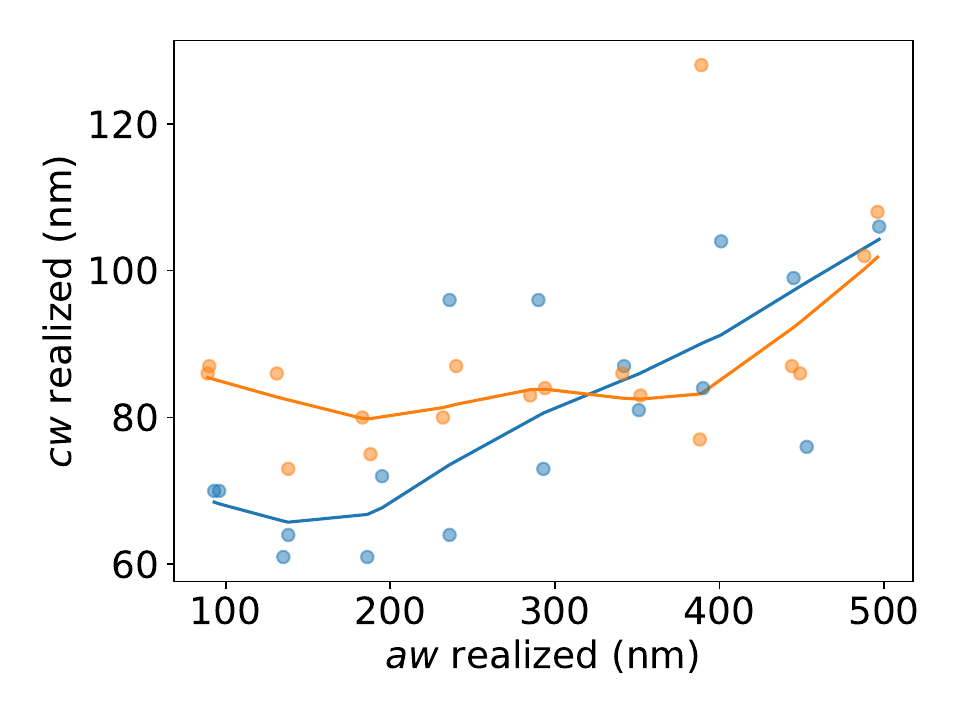}}
\caption{Dependence between realized center width and realized arm width. The lines serve as a guide to the eye, as described in Suppl. Sec.~\ref{sec:SUPP_Fig4}.}
\label{fig:SUPP_cw_real_vs_aw_real}
\end{figure}

\begin{figure}[h]
\centering
\subfigure[~Two-terminal masks.]{\includegraphics[width=0.45\textwidth]{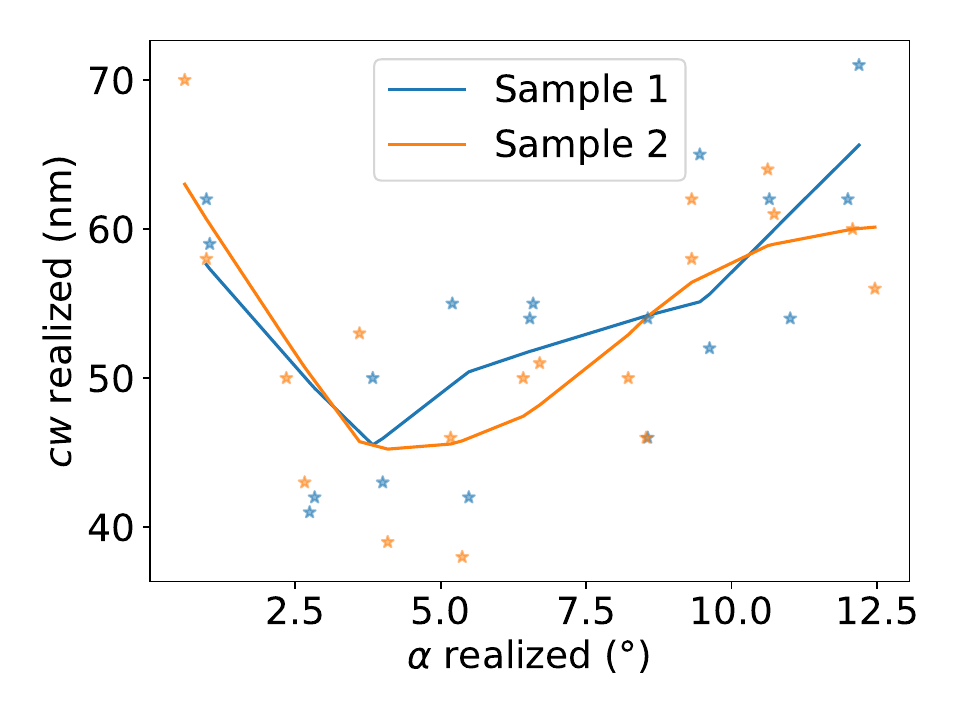}}
\subfigure[~Three-terminal masks.]{\includegraphics[width=0.45\textwidth]{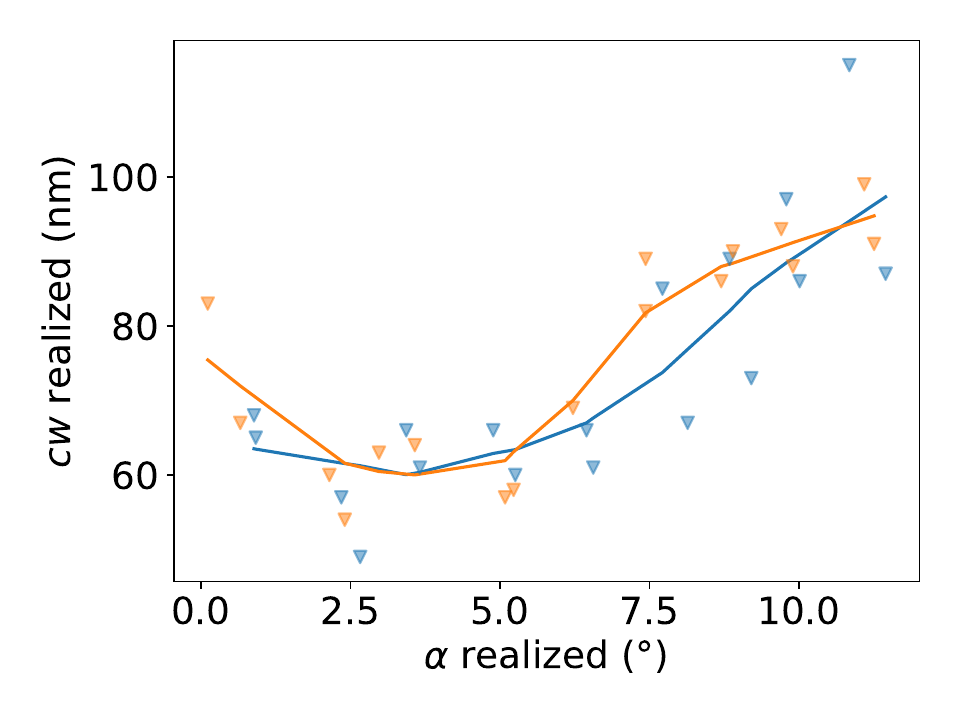}}\\
\subfigure[~Four-terminal masks.]{\includegraphics[width=0.45\textwidth]{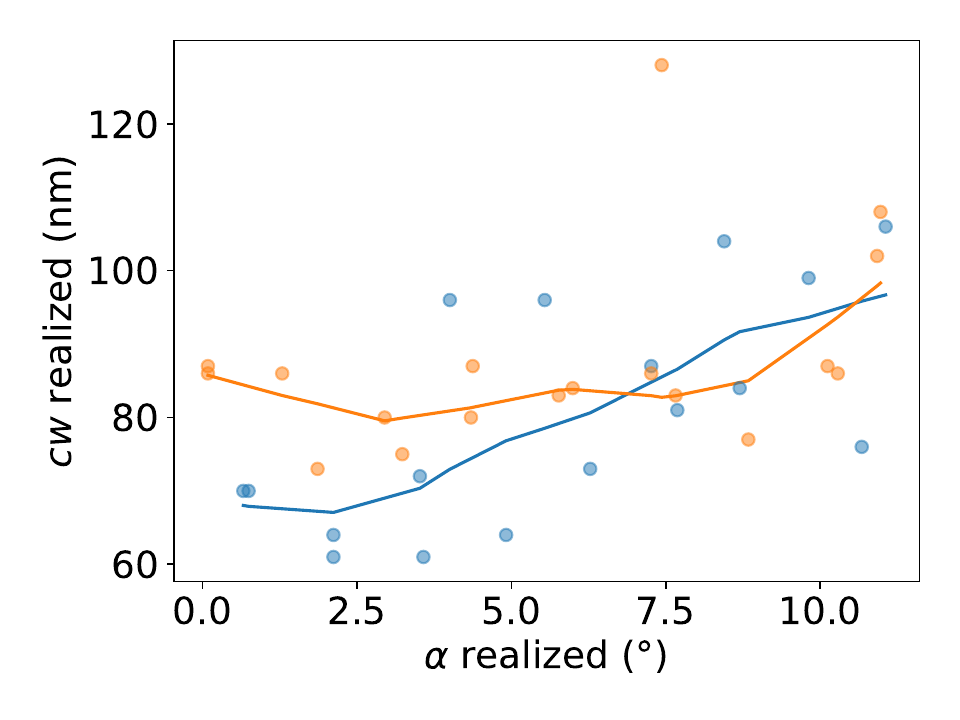}}
\caption{Dependence of the realized center width on the angle of sharpness $\alpha$, assuming an as-designed center length of $cl=1\,\mu$m. The lines serve as a guide to the eye, as described in Suppl. Sec.~\ref{sec:SUPP_Fig4}.}
\label{fig:SUPP_cw_real_vs_alpha}
\end{figure}

Both samples presented in this study were fabricated from the same wafer with the same SiO$_2$/Si$_3$N$_4$ layer stack. They used the same mask designs and were fabricated in the same run. Their measurement results are shown independently in Figs.~\ref{fig:SUPP_Standing_Prob}-\ref{fig:SUPP_cw_real_vs_alpha}. Both samples exhibit comparable behavior.

\end{document}